# Evidence Diet: When Graph Evidence Helps LLMs Rank Competing Candidates

Yuchong Hu†
University of Science and Technology of China
Hefei Anhui China
yuchonghu@mail.ustc.edu.cn

Zhao Zhang†
University of Science and Technology of China
Hefei Anhui China
zhang_zhao@mail.ustc.edu.cn

Ziyao Zhao
University of Science and Technology of China
Hefei Anhui China
ziyao_zhao45@mail.ustc.edu.cn

Haoyu Wang
University of Science and Technology of China
Hefei Anhui China
wanghaoyu@mail.ustc.edu.cn

Jing Xiao*
Ping An Technology (Shenzhen) Co., Ltd., China
Shenzhen Guangdong China
xiaojing661@pingan.com.cn

## ABSTRACT

When several candidates have graph support, does it matter which candidate receives which supporter and path counts? We study how LLM rankers use collaborative evidence on Amazon Books and MovieTV. Swapping user-level evidence degrades ranking, while matched item substitutions have little effect. On fixed clean-trained checkpoints, unrelated catalog text lowers Hit@1 from 44.77% to 38.70%, with 99.9% of rankings still valid. Across all three paired seeds, the NDCG@5 difference between count and path prompts was larger after count training than after path training. Degree matching preserves target paths while giving more rivals support. Across the full degree-matched development pools, adding candidate-level graph records to history improves NDCG@5 by 2.28 points on Books and 2.54 on MovieTV. When the target and a rival both have support, path- and count-trained rankers benefit from retaining counts. On these development queries, adding counts numerically to history scores outperforms presenting them as text. For count-trained rankers, exchanging intact count pairs among supported candidates preserves support and token length, yet lowers NDCG@5 by 4.06 points on 141 Books queries and 3.26 on 104 MovieTV queries. A frozen evaluation on 274 previously unused Books users with supported targets and rivals confirms the attribution effect: correct assignment improves pairwise accuracy by 6.07 percentage points (95% CI [4.18, 8.00]).

## KEYWORDS

LLM recommendation; collaborative evidence; evidence diet; representation compilation.

## 1 Introduction

Web rerankers combine user histories with collaborative evidence [1, 2, 3, 8, 9]. CoRAL retrieves interactions for the prompt [5], MemRec synthesizes collaborative context [10], and LLaRA projects item embeddings into the LLM input space [7]. Prior studies examine context length in temporal and long-context reasoning [4, 12].

We move supporter and path counts among supported candidates while keeping support and the count multiset fixed. This separates identifying supported candidates from ranking among them. It asks which details a ranker still needs once support is known.

(1) Correct count assignment improves target–rival ordering in both domains and on new Books users. (2) Suite I tests source, load, entity binding, and model scale on Books. Suite II varies candidate competition across three pools, compares path- and count-trained rankers across paired seeds, and tests whole-record reassignment with GLM-4-9B-Chat. (3) History scores complement nonlinear scoring in both domains, including on supported Books queries.

## 2 Experimental Design and Protocols

We rank one masked target among ten items per user. Suite I uses connectivity, PageRank (0.85 damping, 30 iterations), and 64-dimensional SVD embeddings. After stripping complete <think>...</think> spans, we extract digit strings. StrictValidRatio is the share of outputs whose extracted IDs form a permutation of 1–10. For H@k/NDCG@k we keep the first occurrence of each valid ID and append missing IDs in ascending order. The Books development graph has 207,759 interactions, 7,377 users and 120,925 items. MovieTV development uses a locally processed copy of InstructRec Amazon Movies/TV [9] (79,737 interactions, 5,649 users, 28,987 items). For the visible Books graph, Suite II removes all 10,929 interactions of 570 users and masks one target interaction for each of the remaining 6,807, leaving 190,023 visible edges. MovieTV masks 5,000 prior targets and one hash-selected edge per user (5,649), leaving 69,088.

**Suite I: Generative Full-Permutation Protocol (§3.1–3.4).** Qwen3 [6] ranks ten Books items. For chronological sequence s, set h=s[:-2]. The first 1,000 users in loader order with len(h)≥7 contribute history h[:-5] to ranking and graph construction, and target h[-5]. Arms share shuffled lists with nine distinct uniform negatives excluding prefix and target; none of the 9,000 appears in its user's complete sequence. We cross 13 inputs, three scales (0.6B/1.7B/4B) and seeds 42–44 (117 settings). At 4B, L0–L3 each train one adapter per seed for one epoch (≤3,000 examples, batch 4, accumulation 4, LoRA rank 16, alpha 32; load definitions in Table 2). Fixed L0/L3 tests vary inference load. Three frozen history-adapted checkpoints each receive six binding probes. A separate experiment crosses flat/hierarchical training and evaluation.

**Suite II: Decoupled Candidate-Logit Protocol (§3.3 and §4).** The original Books-trained Qwen3-4B scores A–J at the final position. We also test paired R/Z-trained rankers and GLM-4-9B-Chat [17] (2,000 Books users × four views, 500 updates, rank-8 fused-QKV LoRA, seed 42). Nine negatives are uniform, exactly target-degree-matched after masking (1,581 Books/684 MovieTV), or top BM25 catalog titles for ≤8 concatenated history titles after excluding each user's full interaction set. History, candidate order and instructions are fixed within each pool. On uniform/BM25 pools, R lists every distinct path from ≤8 displayed history items via other users to each candidate, in sorted order with no path-count cap. The evaluated prompts fit within 4,096 tokens without truncation. R prints NONE for a candidate with no path. N prints the longer OMITTED marker in every slot. Z summarizes exactly the paths shown in R as support b, distinct users c and distinct paths p, omitting identities and incidence. RZ prints both. Paired user bootstrap uses 10,000 draws. Matched (Z−N/Z−Z_mis) and BM25 (Z−N/Z−R) each have four cross-domain contrasts (98.75% CIs). GLM uses 97.5% for two Books contrasts and 95% on MovieTV. Other CIs default to nominal 95%.

The original ranker fits fresh rank-8 LoRA (alpha 16, q_proj/v_proj) on frozen Qwen3-4B. Each of 2,000 Books users supplies four catalog-background-free prompts: two block orders × candidate-specific YES/NO or universally HIDDEN support. We train on final A–J logits (candidate-restricted cross-entropy, no teacher loss; 500 AdamW updates, lr 2e-4, seed 42). R/Z first appear at evaluation; controls include all-zero Z and separately R/Z-trained rankers. Books tuning uses 415 disjoint users; development reuses 1,618 Books and 748 zero-shot MovieTV users. Both new Books cohorts use Amazon Review Data (2018) [16]. For confirmation, we exclude 7,377 prior users and hash-select 2,000 with ≥5 distinct catalog items. Histories require ≥4 distinct catalog items strictly before the target and query the frozen 190,023-edge graph without adding new-user edges. Of 1,770 degree-matched users, 274 (15.5%) have a supported target and rival. Before scoring, we fix eligibility, Z prompts, hashed target slot, order, Z-trained seeds 42–44 and the sole primary endpoint (supported-pair accuracy). We weight all eight label-blind within-support permutations equally, including identity permutations and unchanged inputs; the target need not move. We subtract mean reassigned accuracy from correct accuracy per user and checkpoint, average checkpoint differences within user, then bootstrap users (95% CI).

## 3 Four Diagnostics of Evidence Consumption

### 3.1 Source Decomposition, Overload, and Denoising

**Table 1. Books ranking by evidence source**

| Evidence Configuration | H@1 | H@5 | NDCG@5 |
|---|---|---|---|
| User self summary (structured history) | **45.0** | **76.8** | **61.3** |
| User neighbors (PageRank similar users) | 44.7 | 76.0 | 61.0 |
| History + candidates (baseline) | 44.4 | 75.7 | 60.5 |
| Item neighbors (co-occurrence-based retrieval) | 43.8 | 75.0 | 60.0 |
| History + repeated-history block | 41.8 | 75.6 | 59.4 |
| Item neighbors (SVD-based retrieval) | 40.3 | 70.4 | 55.5 |
| Graph context: user-only / item-only | 40.2 / 39.1 | 71.5 / 70.5 | 56.0 / 55.2 |
| Graph context: joint user + item | 36.6 | 68.7 | 52.8 |
| All signals (unfiltered concatenation) | 31.6 | 67.1 | 49.8 |

*Note.* Qwen3-4B, seeds 42/43/44, 1,000 users per seed (%). Each arm trains its own adapter per seed. The repeated-history block repeats up to eight main-history titles. Nine rows show ten of the 13 input configurations (user-only and item-only share a row). The scale sweep also includes item self-summary, first-order, and second-order evidence.

User summaries lead on Books. The all-signals arm trails history plus candidates by 12.8 H@1 points; joint user–item context trails either block. Co-occurrence beats SVD by 3.5 H@1 and 4.5 NDCG@5 points (Table 1). H@5 falls with neighbor budget (75.57⟶67.10%) and rises with the co-occurrence threshold (57.60⟶67.27%) in separate sweeps (Fig. 1).

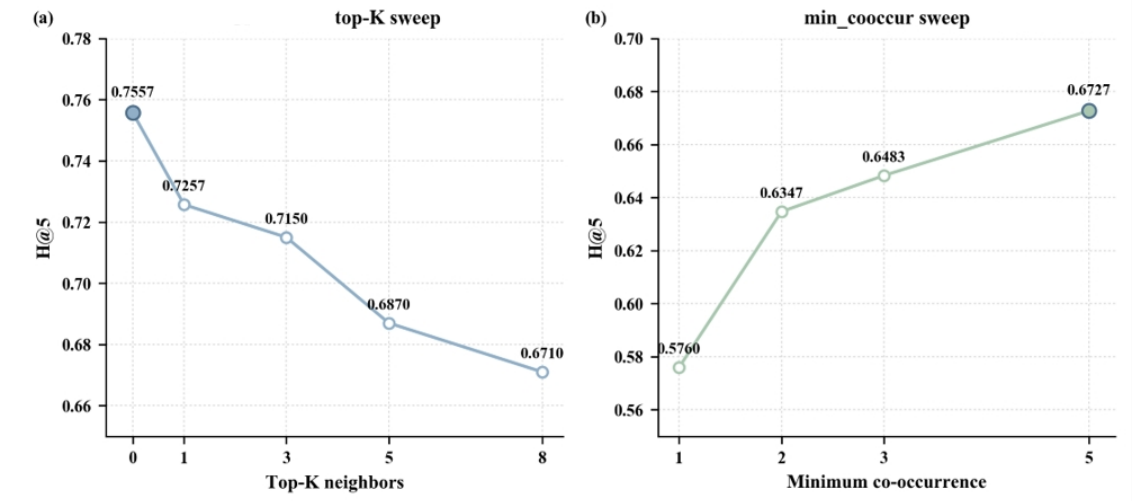

**Figure 1: Books retrieval sweeps (H@5, %). (a) Unfiltered neighbor budget, 0–8. (b) Co-occurrence threshold, 1–5.**

*Note.* Panel (b) is a separate candidate-level retrieval sweep using direct co-reading, two-hop bridges, clustering, and PageRank fallback.

### 3.2 Load-Induced Degradation vs. Output Syntax Validity

L1 retains 100% ranking validity despite lower H@1. L3 trained and evaluated with candidate descriptions reaches 28.3% H@1. With checkpoints fixed, adding unrelated notes to the base prompt lowers H@1 from 44.77% to 38.70% for L0 (6.07-point drop, 95% CI [4.20, 7.97]) and from 18.90% to 15.33% for L3 (3.57-point drop). Loaded L0 rankings remain 99.9% valid (Table 2).

**Table 2. Ranking under prompt load**

| Load Condition | H@1 | NDCG@5 | Valid | Mean Tokens |
|---|---|---|---|---|
| L0: Base prompt (Clean) | 44.7 | 61.4 | 99.9 | 293 |
| L1: Equal-budget unrelated text | 35.4 | 51.8 | 100.0 | 658 |
| L2: Truncate + irrelevant refill | 34.2 | 51.2 | 98.8 | 658 |
| L3: Verbose first-sentence desc. | 28.3 | 46.2 | 99.8 | 657 |
| Fixed L0 (test-time) | 44.77⟶38.70 (−6.07) | 55.9 | 99.9 | 658 |
| Fixed L3 (test-time) | 18.90⟶15.33 (−3.57) | 34.0 | 98.2 | 658 |

*Note.* L0 is the base arm of the load experiment. Table 1 reports separately trained source arms. Suite I, three-seed means on 1,000 users. For fixed checkpoints, H@1 shows clean ⟶ loaded scores; other columns show loaded scores. Differences precede rounding.

### 3.3 When Graph Evidence Distinguishes Competing Candidates

**Entity and format binding.** Cross-user length-matched swaps lower H@1 and target log p (−0.0729 [−0.1146, −0.0319]). Consistent semantic, popularity and dual item-ID replacements yield H@1 estimates within 0.43 points of Real. The oracle uses target co-interactions to select replacement evidence titles, with popularity-matched fallback. Candidates stay fixed and H@1 rises 5.77 points over Real (Table 3). Flat/hierarchical formats retain the evidence body; hierarchical groups by type, and length-matching text varies. Table 3 gives the training × evaluation interaction.

**Paths and counts.** In uniform pools, N leads the listed non-LLM baselines on both metrics, and Z leads the four prompt arms (Table 4A). N uses titles and histories. EASE scores all zero-degree targets at zero (681/1,618 Books, 181/748 MovieTV).

**Table 3. Books entity-binding probes and format interaction**

| Evidence Binding Probe | H@1 | NDCG@5 | log p | H@1 difference vs. Real |
|---|---|---|---|---|
| Real collaborative evidence | 43.00 | 60.06 | -1.8604 | Reference |
| Length-matched cross-user block swap | 40.83 | 58.23 | -1.9333 | -2.17 [-3.33, -1.00] (95% CI) |
| Semantic matched replacement | 43.30 | 59.88 | -1.8474 | +0.30 [−0.67, +1.30] |
| Popularity matched replacement | 42.67 | 59.02 | -1.8461 | −0.33 [−1.63, +0.93] |
| Dual matched (Sem. + Pop.) | 43.43 | 59.80 | -1.8555 | +0.43 [−0.77, +1.63] |
| Target-informed title replacement (oracle) | 48.77 | 63.81 | -1.6645 | +5.77 [+3.90, +7.67] |
| Format interaction[1] | Δ H@1 +3.07 [1.23, 4.93] | | Δ target log p +0.0572 [0.0421, 0.0718] | |

*Note.* Three seeds, 1,000 users per seed. Replacement differences are relative to Real; log p is for the target. Matched and oracle intervals are nominal 95% user-paired estimates. [1] Format interaction: (FF−FH)−(HF−HH), training first (F flat, H hierarchical). Intervals resample users after within-user checkpoint averaging.

On uniform Books, Z beats R by 0.94 NDCG@5 points [0.18, 1.71] despite printing ten explicit records including zero fields (493 vs. 418 tokens). RZ−Z is −0.83 [−1.45, −0.18]. On uniform MovieTV, Z beats R by 0.78 (Table 4A).

Reassigning whole records across candidates (Z_mis) lowers NDCG@5 by 4.38 points on Books and 5.95 on MovieTV, giving the mean ordering Z > N > Z_mis in both uniform pools. Among permutations that transfer support, we contrast the mean logit change of gaining and losing candidates, averaging first within users (Table 4B).

Saturating positive counts at one changes the binary-support-trained ranker's mean uniform-pool NDCG@5 by <0.1 point in either domain. Separately, paired R/Z rankers train on 2,000 Books users (one fixed-layout view, 500 updates). All three seeds show a positive training interaction for Z−R (Table 4B).

**Within-support count attribution.** Full counts improve NDCG@5 over saturated counts under both R and Z training. On Books, Z-trained rankers also achieve higher supported-pair accuracy with full counts than with saturated counts. The training × count interaction remains inconclusive. Separately, reassigning counts preserves support, count pairs, and token length. Correct assignment improves supported-pair accuracy and NDCG@5 in both domains when the target and a rival have support. NDCG@5 gains are smaller across the full matched pools, including on new Books users (Table 4C). For each checkpoint we add $\alpha b + \gamma \log(1+c) + \delta \log(1+p)$ to N logits. Maximizing NDCG@5 on 405/415 exactly matched Books tuning queries over $\alpha \in \{0,1,2,3,4,6\}$, $\gamma,\delta \in \{0,.25,.5,1,2,4\}$ selects (0,2,4), (1,.5,1), (3,2,0) for seeds 42–44, retained on MovieTV.

In supported competition, numerical counts improve on binary bonuses by 3.28 NDCG@5 points on Books [0.33, 6.99]. The MovieTV estimate is +1.63 [−1.46, 4.95]. Numerical counts outperform text Z in both domains (Table 4C). Selecting queries with ≥2 supported candidates yields no clear numerical-over-text advantage. Reassignment also lowers numerical NDCG@5.

**Candidate-pool competition.** Enrichment is target support minus mean negative support, averaged over queries. Four degree-preserving switch chains yield 32 graphs each, rejecting duplicates and forbidden edges. Observed vs. randomized means are 18.28% vs. 4.40% on Books and 28.95% vs. 13.26% on MovieTV (excess 13.88/15.69 points, randomized-mean MCSE 0.06/0.09).

Degree matching retains target paths and raises negative support from 1.45% to 3.81%/8.37% (Books/MovieTV). Z beats N and same-template all-zero Z (Table 4B), and exceeds reassigned whole records by 3.24 [2.47, 4.01]/3.82 [2.53, 5.11] NDCG@5 points.

GLM-4-9B-Chat also favors correct over reassigned records (Books/MovieTV: +1.95 [1.39, 2.53]/+1.69 [0.95, 2.52]). Relative to N, Z gains 1.37 points on Books [0.51, 2.21]. The MovieTV estimate is close to zero (+0.07 [−1.01, 1.12]).

The Books-tuned bonus (N + 3.125b) exceeds Z in uniform and BM25 pools and trails Z in matched pools. All four paired uniform/matched Z−bonus intervals span zero (95%, Table 4B).

**Table 4. Candidate ranking, competition, and count attribution**

| Evaluation Arm / Pool Regime | Information Scope / Compared Arms | Books (Orig/BM25 n=1,618) | MovieTV (Orig/BM25 n=748) |
|---|---|---|---|
| **[Part. A] Popularity ; Title TF-IDF** | Global degree ; lexical cosine (orig) | 24.78 / 36.32 ; 33.37 / 49.64 | 46.52 / 61.32 ; 32.22 / 50.18 |
| User-based CF (UserKNN, K=100) ; EASE [11] | Global interaction graph (orig) | 26.51 / 42.95 ; 22.37 / 34.95 | 33.69 / 50.53 ; 32.35 / 41.85 |
| EASE by visible target degree | Zero / positive strata (each: H@1 / NDCG@5) | 0.29 / 19.13; 38.42 / 46.45 | 1.66 / 21.96; 42.15 / 48.19 |
| Local graph LR (3 features: b, c, p) | Query-local compiled stats (orig) | 26.02 / 42.88 | 35.29 / 52.14 |
| LLM N (History-only baseline) | Ten OMITTED records (423 / 372 toks) | 47.40 / 66.52 | 47.33 / 66.65 |
| LLM R (Raw two-hop paths) | Explicit history-user-item paths (418 / 375 toks) | 49.75 / 68.18 | 52.27 / 69.82 |
| LLM Z (Compiled per-item stats) | Per-item support b, users c, paths p (493 / 442 toks) | 50.43 / 69.12 | 53.48 / 70.60 |
| LLM RZ (Concatenated R + Z) | Raw paths + compiled stats (541 / 498 toks) | 49.44 / 68.29 | 52.14 / 69.79 |
| **[Part. B] Orig. Uniform Pool (NDCG@5)** | N ; R ; Z ; Permuted Z_mis ; Bonus (N+3.125b) | 66.52 ; 68.18 ; 69.12 ; 64.74 ; 69.52 | 66.65 ; 69.82 ; 70.60 ; 64.65 ; 70.68 |
| Deg.-Matched Pool (NDCG@5) | N ; Z ; Permuted Z_mis ; Bonus (n=1,581/684) | 65.09 ; 67.37 ; 64.13 ; 67.16 | 59.11 ; 61.65 ; 57.83 ; 61.19 |
| Same-template zero control | Z − all-zero Z (NDCG@5 diff) | +1.60 [0.62, 2.58] | +2.12 [0.42, 3.89] |
| Title-BM25 Pool (NDCG@5) | N ; R ; Z ; Permuted Z_mis ; Bonus | 17.98 ; 21.94 ; 21.33 ; 18.78 ; 22.10 | 17.90 ; 23.98 ; 24.07 ; 18.61 ; 26.14 |
| Selected contrasts (NDCG@5 diff) | Orig Z-Z_mis ; Deg Z-N [98.75% CI] ; BM25 Z-N | +4.38 ; +2.28 [+1.09,+3.51] ; +3.35 | +5.95 ; +2.54 [+0.51,+4.59] ; +6.17 |
| Original ranker: Count Saturation (Z−Z_sat) | Clamping non-zero c, p > 0 to 1 (n=217/149 altered) | +0.004 [-0.08, +0.09] | +0.076 [-0.02, +0.21] |
| 3-Seed R/Z Train x Eval Interaction | (Z-R)_Z-trained - (Z-R)_R-trained mean (s42/43/44) | +0.70 (+1.66 / +0.07 / +0.39) | +1.11 (+1.54 / +1.01 / +0.77) |
| GLM-4-9B-Chat (Deg.-Matched Pool) | N ; Z ; Z_mis (Z-N / Z-Z_mis NDCG@5 diff) | 66.64; 68.01; 66.05 (+1.37/+1.95) | 65.28; 65.34; 63.65 (+0.07/+1.69) |
| Uniform: score tracking | Gaining − losing support (logits, 95% CI) | 4.61 [4.48,4.73]; n=450 | 4.64 [4.47,4.80]; n=284 |
| Original ranker: supported pairs | Z−N accuracy (pp); post-hoc two-domain 97.5% CI | −2.39 [−8.06,+3.12]; nominal 95% [−7.28,+2.32] | −4.15 [−9.32,+0.46] |
| Original ranker: shared support | N ⟶ Z NDCG@5; Δ [post-hoc 95% CI] | 65.17 ⟶ 78.35; +13.19 [9.00,17.61] | 64.85 ⟶ 79.57; +14.71 [9.67,20.14] |
| Target rank gain (places) | Unsupported rivals / supported rivals / total | +1.057 / −0.057 / +1.000 | +1.144 / −0.058 / +1.087 |
| Single-marker probe | Relative shift (logits); queries with strict reversals | +2.32; 53/64 | +2.41; 53/64 |
| **[C] Z-trained count utility** | Z−Z_sat NDCG; supported competition | +4.24 [1.58,7.24] | +3.32 [0.27,6.82] |
| Count utility: R-trained | Z−Z_sat NDCG; nominal 95% CI | +2.79 [1.08,4.74] | +3.26 [0.88,5.76] |
| Z-trained supported-pair gain | Z−Z_sat net correction (pp) | +4.37 [0.84,8.06] | +3.49 [−0.18,7.96] |
| Training × count interaction | Z-trained minus R-trained: NDCG / pair gain | +1.45 [−0.49,3.45] / +1.05 [−2.57,4.65] | +0.06 [−1.98,2.15] / −0.37 [−3.51,2.86] |
| Z-trained: within-support count reassignment | True−reassigned counts: NDCG | +4.06 [1.72,6.81] (3.84/4.25/4.09) | +3.26 [0.48,6.50] (2.38/3.30/4.09) |
| Z-trained attribution: supported pairs | True−reassigned: pair accuracy (pp) | +5.14 [2.17,8.69] (4.99/5.37/5.05) | +3.55 [0.60,6.88] (1.61/3.87/5.18) |
| Numerical − text: supported target and rival | Numerical − text Z: NDCG@5 points (99.17% CI) | +5.38 [2.45,8.50] | +3.88 [0.67,7.13] |
| Numerical − text: ≥2 supported | Observable before ranking (97.5% CI) | +0.86 [−1.61,3.39] | −0.42 [−3.23,2.32] |
| Numerical − text: full pool | All degree-matched queries | ≈+0.26 | ≈−0.10 |
| Development: numerical count readout | Correct − reassigned counts: NDCG@5 | +5.84 | +5.35 |
| Full matched cohorts | Text Z: correct − reassigned counts, NDCG@5 | Dev +0.36 (1,581); new +0.56 (1,770) | Dev +0.53 (684) |
| New-user confirmation | Supported-pair accuracy (%): reassigned ⟶ correct | 66.87 ⟶ 72.95; Δ +6.07 [4.18, 8.00] | — |

*Note.* A: H@1/NDCG@5 (%); uniform Z−R/RZ−Z, nominal 95% CIs. Matched/BM25 primary contrasts: 98.75% (§2). B: training × evaluation by seed. C: supported target and rival unless specified (Books 141/1,581, 8.9%; MovieTV 104/684, 15.2%). User bootstrap conditions on fixed checkpoints; multi-checkpoint metrics or contrasts are averaged within user. Attribution uses eight label-blind permutations (parentheses: seeds 42/43/44). Differences precede rounding. All-zero, count utility, training × count, and reassignment each form a separate four-test family (designated contrasts, 98.75%). Subgroup numerical–text/count–binary and shuffled Z–N form one six-test family (99.17%). The new Books primary endpoint uses a 95% CI.

A binary bonus preserves within-support order under count reassignment. With the original ranker, supported targets gain rank against unsupported rivals and lose ground slightly against supported rivals in both domains (Table 4B). Against supported rivals, Z−N pair accuracy changes by −2.39 points on Books and −4.15 on MovieTV; both two-domain-adjusted intervals span zero.

Table 4C separately compares full and saturated counts in Z-trained rankers. Target support contributes positively to Z–N, support confined to negatives negatively. Users receive equal weight for single-target NDCG@5 and supported-pair accuracy (fraction of supported rivals below the target, candidate-index tie-break).

All nine BM25 negatives lexically outrank the target for 94.75% of Books and 92.25% of MovieTV queries (1,533/1,618; 690/748). In retrieval order, Books/MovieTV Z–N gains are 3.35 [2.04, 4.69]/6.17 [4.25, 8.11]; Z–R is −0.61 [−1.60, 0.36]/+0.09 [−1.41, 1.63]. Across four label-blind presentation orders, each candidate moves with its record. Z–N gains remain 3.31 [2.30, 4.37]/5.78 [4.26, 7.37].

One label-blind slot changes from (0,0,0) to (1,1,1), leaving nine zero records (Table 4B). Center the ten logit differences (one-marker minus all-zero) within query to obtain D. For marked slot m, $v = e_m - \mathbb{1}/10$. Least-squares shifts explain $1-\Sigma\|D-gv\|^2/\Sigma\|D\|^2$ of change energy: 72.45%/76.00% with shared g per domain, 88.63%/89.19% with query-specific g (Books/MovieTV). Among 36 unmarked pairs per query, 53/64 queries per domain have strict reversals (132/159 pairs, ties excluded).

**History–count conflicts.** We screen 100,000 additional Books users, excluding 7,377 prior users and 2,000 confirmation selections. For the predesignated seed-42 Z-trained checkpoint, history-only N and frozen count scores strictly disagree on 6,831 supported target–rival pairs (4,593 users). On these pairs, averaging within users, then across users, gives 41.74% (N), 55.16% (text Z) and 58.26% (numerical). Numerical–text is +3.10 points [0.87, 5.36] (97.5%, Bonferroni over Z–N and Z–numerical).

### 3.4 Model Scale and Ranking-Format Validity

Fewer than half the configurations at either smaller scale reach 90% strict validity (Table 5). Ranking metrics include repaired outputs.

**Table 5. Ranking quality and output validity by model scale**

| Model Scale | H@1 | H@3 | H@5 | NDCG@5 | Configs ≥90% valid |
|---|---|---|---|---|---|
| Qwen3-0.6B | 14.4 | 38.3 | 57.7 | 36.1 | 10 / 39 (25.6%) |
| Qwen3-1.7B | 22.1 | 45.6 | 62.9 | 42.6 | 17 / 39 (43.6%) |
| Qwen3-4B | 36.0 | 55.3 | 68.9 | 52.7 | 39 / 39 (100.0%) |

*Note.* Books, 13 sources × three seeds per scale. At 0.6B and 1.7B, metrics average all 39 configurations using the §2 repair rule. The pass rate counts configurations reaching 90% strict validity.

## 4 Where Complementary LLM Gains Occur

We fuse scores as $F=z(S)+\lambda z(N)$, standardizing across the ten candidates within each query (population SD floored at $10^{-6}$). LR and scikit-learn's HistGradientBoostingClassifier (HGB) extend the three-feature LR (Table 4A) with title TF-IDF dotted with the mean history vector and visible item degree, alongside support and log supporter/path counts. Disjoint train/tuning/assessment sizes are 2,000/415/1,618 (Books) and 2,000/225/748 (MovieTV). LR tests $C\in\{0.1,1,10\}$. HGB specifies seeds 42–44 and 12 settings (leaves 7/15/31 × rates .05/.1 × L2 0/1), selecting by H@1 then NDCG@5. Books tuning selects $\lambda$=1 (LR) and .5 (HGB) from {0,.25,.5,1,2}; both transfer to MovieTV.

We remove visible item degree from the five-feature LR in Table 6 and refit on the remaining four features. Adding N raises H@1 from 41.78% to 47.22% on Books (+5.44 points [3.83, 7.05]) and from 45.99% to 51.60% on MovieTV (+5.61 [3.21, 8.02]).

History scores complement HGB in both domains. Supported Books queries also gain; all net MovieTV top-1 gains come from zero-support queries (Table 6).

LR fusion improves overall NDCG@5 in both domains. Supported queries gain little at top 1, though Books NDCG@5 improves (Table 6).

**Table 6. Structural and history-score fusion by support group**

| Domain & Subgroup | Users | Graph + title LR (S, 5 features) | LLM N | Fusion F | + / - (Net) | ΔH@1 / ΔNDCG@5 (points) |
|---|---|---|---|---|---|---|
| Books: All users | 1,618 | 51.11 / 65.43 | 47.40 / 66.52 | 54.45 / 71.05 | 159 / 105 (+54) | +3.34 [+1.36,+5.25] / +5.62 [+4.27,+6.95] |
| Any supported candidate | 450 | 67.11 / 78.32 | 54.67 / 71.08 | 67.56 / 81.30 | 22 / 20 (+2) | +0.44 [-2.44,+3.12] / +2.98 |
| No supported candidate | 1,168 | 44.95 / 60.47 | 44.61 / 64.76 | 49.40 / 67.10 | 137 / 85 (+52) | +4.45 [+1.97,+6.93] / +6.64 |
| MovieTV: All users | 748 | 56.28 / 70.27 | 47.33 / 66.65 | 58.69 / 75.24 | 64 / 46 (+18) | +2.41 [-0.27,+5.21] / +4.97 [+3.22,+6.72] |
| Any supported candidate | 284 | 80.63 / 89.36 | 63.73 / 77.97 | 79.93 / 89.95 | 12 / 14 (-2) | -0.70 [-4.23,+2.82] / +0.58 |
| No supported candidate | 464 | 41.38 / 58.59 | 37.28 / 59.72 | 45.69 / 66.24 | 52 / 32 (+20) | +4.31 [+0.43,+8.19] / +7.65 |
| **Same-feature HGB control** | **Users** | **HGB S** | **LLM N** | **HGB + N** | **+ / – (Net)** | **Diff H@1 / NDCG@5** |
| Books: All users | 1,618 | 61.93 / 77.97 | 47.40 / 66.52 | 67.92 / 82.11 | 155 / 58 (+97) | +6.00 [4.20,7.79] / +4.14 [3.24,5.04] |
| MovieTV: All users | 748 | 56.68 / 72.25 | 47.33 / 66.65 | 60.43 / 76.12 | 42 / 14 (+28) | +3.74 [1.87,5.75] / +3.87 [2.48,5.21] |
| Books: Zero support | 1,168 | 58.13[1] | 44.61 / 64.76 | 64.73[1] | 122 / 45 (+77) | +6.59 [3.94,9.33] / +4.68 [3.25,6.11] |
| MovieTV: Zero support | 464 | 40.95[1] | 37.28 / 59.72 | 46.98[1] | 36 / 8 (+28) | +6.03 [2.59,9.70] / +6.09 [3.47,8.72] |
| Books: Any-support | 450 | 71.78[1] | 54.67[1] | 76.22[1] | 33 / 13 (+20) | +4.44 [1.56,7.33] / +2.74 [1.42,4.07] |
| MovieTV: Any-support | 284 | 82.39[1] | 63.73[1] | 82.39[1] | 6 / 6 (0) | 0.00 [−2.46,2.46] / +0.25 [−1.00,1.46] |

*Note.* Suite II. S, N, and F give H@1 / NDCG@5 (%). +/− counts top-1 corrections/regressions of F vs. S. Diff gives paired F–S gains in points. For LR subgroups, brackets give H@1 intervals; NDCG@5 entries are point estimates. [1] H@1 only. HGB uses nominal 95% intervals overall and on supported queries, and four-comparison-adjusted 98.75% intervals on zero-support queries. Differences precede rounding.

Zero-support queries account for 52 of 54 net LR gains on Books (72.2% of queries). On MovieTV, they account for 62.0% of queries and add 20 net hits; supported queries lose two, leaving 18 overall (Table 6). Zero-minus-supported gains are +4.01 [0.21, 7.79]/+3.65 [1.07, 6.23] on Books and +5.01/+7.07 [3.94, 10.22] on MovieTV (H@1/NDCG@5). Restricting offline LR fusion to zero-support queries skips 27.8%/38.0% of LLM calls at a cost of 0.83/0.22 NDCG@5 points (70.22 vs. 71.05; 75.02 vs. 75.24).

## 5 Related Work and Conclusion

CoRAL and MemRec retrieve collaborative context [5, 10], LLaRA aligns behavioral embeddings [7], and GLRec prompts with paths [13]. GARDRec grounds ranking decisions in graph evidence [14]. HopQA tests shortest-hop recovery [15]. Holding support and the count multiset fixed, correct assignment improves target–rival ordering in both development domains and in a frozen test on 274 new Books users (+6.07 percentage points in supported-pair accuracy). Across all degree-matched development queries, the same count-assignment intervention improves NDCG@5 by 0.36 points on Books and 0.53 on MovieTV (Table 4C). Degree matching preserves target paths but increases rival support. Correct evidence also helps under BM25 retrieval. When targets and rivals share support, both path- and count-trained rankers benefit from counts, and numerical injection outperforms text on the same history scores. History scores complement nonlinear scoring even on supported Books queries.

## REFERENCES

[1] Shijie Geng, Shuchang Liu, Zuohui Fu, Yingqiang Ge, and Yongfeng Zhang. 2022. Recommendation as Language Processing (P5). In RecSys '22. 299-315.

[2] Keqin Bao, Jizhi Zhang, Yang Zhang, Wenjie Wang, Fuli Feng, and Xiangnan He. 2023. TALLRec: An Effective and Efficient Tuning Framework to Align LLM with Recommendation. In RecSys '23. 1007-1014.

[3] Xiangnan He, Kuan Deng, Xiang Wang, Yan Li, YongDong Zhang, and Meng Wang. 2020. LightGCN: Simplifying and Powering Graph Convolution Network for Recommendation. In SIGIR '20. 639-648.

[4] Nelson F. Liu, Kevin Lin, John Hewitt, Ashwin Paranjape, Michele Bevilacqua, Fabio Petroni, and Percy Liang. 2024. Lost in the Middle: How Language Models Use Long Contexts. TACL 12 (2024), 157-173.

[5] Junda Wu, Cheng-Chun Chang, Tong Yu, Zhankui He, Jianing Wang, Yupeng Hou, and Julian McAuley. 2024. CoRAL: Collaborative Retrieval-Augmented Large Language Models Improve Long-tail Recommendation. In KDD '24. 3391-3401.

[6] An Yang et al. 2025. Qwen3 Technical Report. arXiv:2505.09388 (2025).

[7] Jiayi Liao, Sihang Li, Zhengyi Yang, Jiancan Wu, Yancheng Yuan, Xiang Wang, and Xiangnan He. 2024. LLaRA: Large Language-Recommendation Assistant. In SIGIR '24. 1785-1795.

[8] Junjie Zhang, Yupeng Hou, R. Xie, W. Sun, J. McAuley, W. X. Zhao, L. Lin, and J.-R. Wen. 2024. AgentCF: Collaborative Learning with Autonomous Language Agents for Recommender Systems. In WWW '24. 3679-3689.

[9] Wujiang Xu et al. 2025. iAgent: LLM Agent as a Shield between User and Recommender Systems. In Findings of ACL '25. 18056-18084.

[10] Weixin Chen et al. 2026. MemRec: Collaborative Memory-Augmented Agentic Recommender System. In ACL '26 (Volume 1: Long Papers), 44515–44544.

[11] Harald Steck. 2019. Embarrassingly Shallow Autoencoders for Sparse Data (EASE). In WWW '19. 3251-3257.

[12] Yanan Cao et al. 2026. Is More Context Always Better? Examining LLM Reasoning Capability for Time Interval Prediction. In Proceedings of the ACM Web Conference 2026, 8521–8524.

[13] Likang Wu, Zhaopeng Qiu, Zhi Zheng, Hengshu Zhu, and Enhong Chen. 2024. Exploring Large Language Model for Graph Data Understanding in Online Job Recommendations. In AAAI '24. doi:10.1609/aaai.v38i8.28769.

[14] Yong Wang, Hongliang Sun, Jinlan Liu, Hua Zhang, Dianbo Sui, Dianhui Chu, and Zhiying Tu. 2026. GARDRec: Decision-Level Graph Grounding for Large Language Model Recommendation. arXiv:2608.00669.

[15] Xiaoyu Guo, Pengcheng Chen, Jiong Yu, Yi Lu, Yaohua Wang, and Ziyang Li. 2026. Graph Evidence Is Not Enough: Diagnosing Native Decoder Use in Graph-Augmented LLMs. arXiv:2608.30437.

[16] Jianmo Ni, Jiacheng Li, and Julian McAuley. 2019. Justifying Recommendations using Distantly-Labeled Reviews and Fine-Grained Aspects. In EMNLP-IJCNLP, 188–197.

[17] GLM Team et al. 2024. ChatGLM: A Family of Large Language Models from GLM-130B to GLM-4 All Tools. arXiv:2406.12793.